\documentclass[aps,pra,twocolumn,nofootinbib,10pt]{revtex4-1}

\usepackage{natbib}
\usepackage{amsmath,amssymb,bm}
\usepackage{dsfont}
\usepackage{graphics}

\newcommand{\tr}{\text{tr}}

\newcommand{\Dcirc}{{\cal D}^{\circ}}
\newcommand{\dbar}{d\hspace*{-0.08em}\bar{}\hspace*{0.1em}}

\begin{document}

\title{Spatiotemproal effects on squeezing measurements}
\author{Filippus S. \surname{Roux}}
\email{froux@nmisa.org}
\affiliation{National Metrology Institute of South Africa, Meiring Naud{\'e} Road, Brummeria 0040, Pretoria, South Africa}

\begin{abstract}
The role of the spatiotemporal degrees of freedom in the preparation and observation of squeezed photonic states, produced by parametric down-conversion, is investigated. The analysis is done with the aid of a functional approach under the semi-classical approximation and the thin-crystal approximation. It is found that the squeezed state loses its minimum uncertainty property as the efficiency of down-conversion is increased, in a way that depends on the conditions of the homodyne measurements with which the amount of squeezing is determined.
\end{abstract}

\maketitle

\section{\label{intro}Introduction}

Squeezed vacuum states \cite{sqlight,lvovsqu} are among the most useful quantum states. They are used to make measurements that are more accurate than the standard quantum limit (the shot-noise limit) would allow. For this reason, they are used in various applications, ranging from quantum computing to gravitational wave detection \cite{kolobov,braunstein,weedbrook,contvar1,contvar2,squreview,ligosqu}.

In quantum optics, squeezed vacuum states can be produced with the aid of nonlinear optical processes, such as parametric down-conversion \cite{mandelspdc}. Although challenging to analyze, parametric down-conversion has been studied for decades and much progress has been made in its analytical representation \cite{arnaut,bennink,law,miatto3,gattispdc,pdcspatcor,oamsqu3,horoshko}.

A squeezed vacuum state can be represented by a squeezing operator applied to the vacuum. Without the spatiotemporal degrees of freedom, the squeezing operator is parameterized by a single parameter, called the {\em squeezing parameter}. The inclusion of the spatiotemporal degrees of freedom causes the squeezed vacuum states to become a complex combination of {\em squeezing eigenmodes}, each with its own squeezing parameter \cite{bennink}.

To incorporate the spatiotemporal degrees of freedom in a down-converted squeezed vacuum states, the squeezing operator is often expressed in terms of a set of ladder operators (or quadrature operators) in which each operator pair is associated with one of the eigenmodes \cite{bennink,oamsqu3,horoshko}. It leads to a symplectic formalism where the squeezing operator is represented as a Gaussian operator --- an exponential with an argument that is second order in ladder (or quadrature) operators \cite{braunstein,weedbrook,contvar1,contvar2}.

The versatility of the symplectic formalism is demonstrated by its use in numerous analyses of the parametric down-conversion process and its applications. It represents the kernels of the down-conversion process as matrices that can be diagonalized with a Bloch-Messiah reduction \cite{blochmessiah}. Such a representation assumes a discrete finite dimensional system. The down-conversion process can also be represented in terms of a Magnus expansion \cite{magnus,magnusrev}. The matrix representation usually involves only the first-order in the Magnus expansion \cite{horoshko}.

Moreover, the matrix representation associated with the Bloch-Messiah reduction assumes knowledge of the eigenmodes (or Schmidt basis) of the kernels. However, the exact nature of these eigenmodes are not known and has so far only been obtained approximately by employing simplifying assumptions \cite{miatto3,law}.

The applicability of the symplectic formalism and the Bloch-Messiah reduction in the analysis of the parametric down-conversion process is limited to those cases where the pump can be represented as a classical field, variously called the {\em semiclassical approximation}, the {\em parametric approximation}, the {\em undepleted pump approximation}, among others. The justification for this approximation is that the efficiency of the process is low enough that the pump remains unaffected (undepleted).

Recently, a functional formalism was used to analyze the down-conversion process, incorporating all spatiotemporal degrees of freedom \cite{nosemi}. The functional formalism operates directly with the kernels of the down-conversion process, and does not need to represent them as finite dimensional matrices. It avoids the need for knowledge of the eigenmodes by expressing any final result (such as what would be obtained in a measurement) directly in terms of the kernels. The functional formalism also allows calculations beyond the semiclassical approximation, with the aid of a perturbative process. It reproduces the expressions of the semiclassical part of the solution (zeroth order in the perturbative expansion) as a functional version of the Magnus expansion.

However, without knowledge of the eigenmodes, the question of the optimal choice for the mode of a local oscillator to be used in homodyne measurements is left open. Since the different eigenmodes are associated with different amounts of squeezing, the optimal amount of squeezing should be obtained when the local oscillator has the same mode as that of the associated eigenmode.

In this paper, we consider the effect of the spatiotemporal properties of the local oscillator directly in terms of the kernels. We perform the calculations with the aid of the Wigner functional formalism \cite{stquad,stquaderr} to all orders in the Magnus expansion of the down-converted state, under the so-called {\em thin-crystal} approximation. The latter is well satisfied in most experimental conditions, especially when the spatial degrees of freedom play an important role \cite{law}. The effect of higher orders of the Magnus expansion has previously been investigated in four-wave mixing, for the temporal degrees of freedom only \cite{sipemagnus1,sipemagnus2}.

Numerous spatiotemporal properties can play a role in homodyne measurements. Here, we consider the effects of the mode size for a Gaussian profile and the temporal bandwidth of the local oscillator. We also consider the effect of the azimuthal index, if the local oscillator has a Laguerre-Gauss petal mode profile with zero radial index. The Laguerre-Gauss modal basis is an example of an orbital angular momentum (OAM) basis. Recently, the incorporation of OAM modes into squeezed states has attracted much attention \cite{oamsqu1,oamsqu2,oamsqu3,cvoamappl,oamsquappl}. Our interest is to determine the effect of the azimuthal index (OAM) on homodyning measurements.

Contrary to our previous work \cite{nosemi}, we perform the calculations for the current investigation under the semi-classical approximation only. The validity for this approximation in the current situation is however not based on an assumed low efficiency for the down-conversion process. It is justified simply on the assumption that the pump remains a coherent state during the process. As such, the relative size of the minimum uncertainty area of the pump on the functional phase space ensures that it can be adequately approximated with the classical field that parametrizes the coherent state of the pump.

Here, we apply a weak monochromatic assumption: we assume that the pump and the local oscillator both have narrow spectra, but with finite bandwidths. It allows us to determine the bandwidth of the different terms in the Magnus expansion of the down-converted state, which plays an important role in the results that we obtain.

\section{General approach}

In order to determine how successful a strongly driven down-conversion process is at producing a squeezed vacuum state, one can measure the amount of squeezing of the state, using a homodyne measurement. Here, we calculate what would be obtained from such a measurement of the squeezing parameter. It is done by computing the uncertainty (variance) in the measurement of the quadratures of the down-converted state. The variance in the quadrature is given by
\begin{equation}
\sigma_q^2 \equiv \langle \Delta \hat{q}_{\theta}^2 \rangle = \langle \hat{q}_{\theta}^2 \rangle-\langle \hat{q}_{\theta} \rangle^2
 = \tr\{\hat{\rho} \hat{q}_{\theta}^2\} - \tr\{\hat{\rho} \hat{q}_{\theta}\}^2 ,
\label{varqdef}
\end{equation}
where the generalized quadrature operator is defined as
\begin{equation}
\hat{q}_{\theta} \equiv \hat{q}\cos\theta - \hat{p}\sin\theta ,
\label{genqdef}
\end{equation}
with $\theta$ representing the orientation in phase space.

In terms of the spatiotemporal degrees of freedom, a homodyne measurement selects only a specific spatiotemporal mode given by the modulus of the mode of the local oscillator. The phase of the local oscillator determines the orientation of the generalized quadrature operator in phase space. The observable for the homodyne measurement is therefore represented by
\begin{equation}
\hat{h}[\gamma] = \sqrt{2} \int |\gamma(\mathbf{k})| \hat{q}_{\theta}(\mathbf{k})\ \dbar k ,
\label{defdynop}
\end{equation}
where $\gamma(\mathbf{k})=|\gamma(\mathbf{k})|\exp(i\theta)$\footnote{Here, it is assumed that the phase is a global constant. In general it can also be a function of the wave vector $\theta(\mathbf{k})$.} represents the parameter function for the coherent state of the local oscillator, $\mathbf{k}$ is the three-dimensional wave vector, and
\begin{equation}
\dbar k\equiv \frac{\text{d}^3 k}{(2\pi)^3 \omega} .
\label{dkmaat}
\end{equation}

We'll use a Wigner functional formalism \cite{entpdc,nosemi} to perform the computations. The operator for a homodyne measurement is represented by the Wigner functional
\begin{equation}
W_{\hat{h}}[\alpha;\gamma] = \gamma^*\diamond\alpha+\alpha^*\diamond\gamma ,
\end{equation}
where the $\diamond$-contraction is defined as \cite{entpdc,nosemi}
\begin{equation}
\alpha_1^*\diamond\alpha_2 \equiv \int \alpha_1^*(\mathbf{k}) \alpha_2(\mathbf{k})\ \dbar k ,
\end{equation}
and
\begin{equation}
\alpha(\mathbf{k}) = \frac{1}{\sqrt{2}} \left[q(\mathbf{k}) + i p(\mathbf{k})\right] ,
\label{alphadef}
\end{equation}
is a complex-valued function that parametrizes the functional phase space, with $q(\mathbf{k})$ and $p(\mathbf{k})$ being the eigenvalue functions of the quadrature operators \cite{stquad,stquaderr}.

To alleviate the calculation of the expectation values, we define a generating functional given by
\begin{equation}
\mathcal{W}[\gamma](\eta) = \int \exp\left(\eta W_{\hat{h}}[\alpha;\gamma]\right) W_{\hat{\rho}}[\alpha]\ \Dcirc[\alpha] ,
\label{genverwq}
\end{equation}
where $W_{\hat{\rho}}[\alpha]$ is the Wigner functional of the state, $\eta$ is a generating parameter, and
\begin{equation}
\Dcirc[\alpha] \equiv \mathcal{D}[q]\ \mathcal{D}\left[\frac{p}{2\pi}\right] .
\label{alfamaat}
\end{equation}
The expectation values are obtained by computing
\begin{align}
\begin{split}
\langle \hat{h}[\gamma] \rangle = \sqrt{2} \gamma_0 \langle \hat{q}_{\theta} \rangle
 = & \left. \partial_{\eta} \mathcal{W}[\gamma](\eta) \right|_{\eta=0} , \\
\langle \hat{h}^2[\gamma] \rangle = 2 \gamma_0^2 \langle \hat{q}_{\theta}^2 \rangle
 = & \left. \partial_{\eta}^2 \mathcal{W}[\gamma](\eta) \right|_{\eta=0} ,
\end{split}
\label{verwq}
\end{align}
where
\begin{equation}
\gamma_0^2=\|\gamma(\mathbf{k})\|^2 \equiv \int |\gamma(\mathbf{k})|^2\ \dbar k .
\end{equation}

\section{Squeezed vacuum state}

The Wigner functional for the squeezed vacuum state produced by parametric down-conversion is given by \cite{nosemi}
\begin{align}
W_{\hat{\rho}}[\alpha] = & \mathcal{N}_0
\exp\left(-2\alpha^*\diamond A\diamond\alpha-\alpha\diamond B\diamond\alpha \right. \nonumber \\
& \left. -\alpha^*\diamond B^*\diamond\alpha^*\right) ,
\label{ansatz}
\end{align}
where $\mathcal{N}_0$ is the normalization constant for the Wigner functional of a pure Gaussian state, and $A$ and $B$ are semi-classical kernel functions associated with the parametric down-conversion process.

When we substitute Eq.~(\ref{ansatz}) into Eq.~(\ref{genverwq}) and evaluate the functional integral over $\alpha$, we obtain
\begin{align}
\mathcal{W}(\eta) = & \mathcal{N}_0 \int \exp\left[-2\alpha^*\diamond A\diamond\alpha
-\alpha\diamond B^*\diamond\alpha \right. \nonumber \\
& \left. -\alpha^*\diamond B\diamond\alpha^* +\eta\left(\gamma^*\diamond\alpha+\alpha^*\diamond\gamma\right) \right]\ \Dcirc[\alpha] \nonumber \\
= & \exp\left(\frac{\eta^2}{2}\gamma^*\diamond A\diamond\gamma
 -i \frac{\eta^2}{4}\gamma\diamond A^*\diamond B^*\diamond A^{-1}\diamond\gamma \right. \nonumber \\
& \left. +i \frac{\eta^2}{4}\gamma^*\diamond A^{-1}\diamond B\diamond A^*\diamond\gamma^* \right) .
\label{svgenverwq}
\end{align}

Computing the expectation values, as in Eq.~(\ref{verwq}), we obtain $\langle\hat{h}\rangle=0$ because the squeezed vacuum state is centered at the origin. The expression for $\langle\hat{h}^2\rangle$ is
\begin{align}
\langle\hat{h}^2\rangle = & \left. \partial_{\eta}^2 \mathcal{W}(\eta) \right|_{\eta=0} \nonumber \\
= & \gamma^*\diamond A\diamond\gamma
-i \tfrac{1}{2}\gamma\diamond A^*\diamond B^*\diamond A^{-1}\diamond\gamma \nonumber \\
& +i \tfrac{1}{2}\gamma^*\diamond A^{-1}\diamond B\diamond A^*\diamond\gamma^* \nonumber \\
 = & \gamma^*\diamond A\diamond\gamma -i \tfrac{1}{2}\gamma\diamond B^*\diamond\gamma
 +i \tfrac{1}{2}\gamma^*\diamond B\diamond\gamma^* ,
\label{qqverw}
\end{align}
where the final result follows under the assumption that $A$ and $B$ commute and that $A$ is real-valued.

\begin{widetext}
\section{Semi-classical kernel functions}

Under the semi-classical approximation, the Wigner functional of the down-converted state can be represented in terms of a functional Magnus expansion \cite{horoshko,nosemi}, in which the bilinear kernels in the exponent are given by
\begin{align}
\begin{split}
A = & \mathbf{1} + \int_0^L \int_0^{z_1} \mathcal{Z}\left\{ H^*(z_1)\diamond H(z_2) \right\}\ \text{d}z_2\ \text{d}z_1 \\
& + \int_0^L \int_0^{z_1} \int_0^{z_2} \int_0^{z_3} \mathcal{Z}\left\{ H^*(z_1)\diamond H(z_2)\diamond H^*(z_3)\diamond H(z_4)
\right\}\ \text{d}z_4\ \text{d}z_3\ \text{d}z_2\ \text{d}z_1 + ... , \\
B = & \int_0^{L} H(z_1)\ \text{d}z_1 + \int_0^L \int_0^{z_1} \int_0^{z_2}
\mathcal{Z}\left\{ H(z_1)\diamond H^*(z_2)\diamond H(z_3) \right\}\ \text{d}z_3\ \text{d}z_2\ \text{d}z_1 \\
& + \int_0^L  \int_0^{z_1} \int_0^{z_2} \int_0^{z_3} \int_0^{z_4}
\mathcal{Z}\left\{ H(z_1)\diamond H^*(z_2)\diamond H(z_3) \diamond H^*(z_4) \diamond H(z_5) \right\} \\
& \times\ \text{d}z_5\ \text{d}z_4\ \text{d}z_3\ \text{d}z_2\ \text{d}z_1 + ...\ .
\end{split}
\label{defzab}
\end{align}
Here, $L$ is the length of the nonlinear crystal, $H(z)$ is the bilinear vertex kernel that is obtained by contracting the classical parameter function of the pump on the vertex for the down-conversion process, and $\mathcal{Z}\{\cdot\}$ represents a symmetrization operation, which is recursively defined by
\begin{equation}
\mathcal{Z}\left\{ f_1(z_1)\diamond ...\diamond f_n(z_n) \right\} =
\tfrac{1}{2} f_1(z_1)\diamond \mathcal{Z}\left\{ f_2(z_2)\diamond ...\diamond f_n(z_n) \right\}
+\tfrac{1}{2} \mathcal{Z}\left\{ f_1(z_2)\diamond ...\diamond f_{n-1}(z_n) \right\}\diamond f_n(z_1) ,
\end{equation}
with $\mathcal{Z}\{f_1(z_1)\}=f_1(z_1)$. The identity $\mathbf{1}$ is defined so that $\int \mathbf{1}\ \dbar k = 1$.
\end{widetext}

The monochromatic parameter function of the pump is assumed to be
\begin{equation}
\zeta(\mathbf{k}) = \zeta_0 w_{\text{p}} \sqrt{\frac{2\pi \omega}{c}} S(\omega-\omega_{\text{p}};\delta\omega_{\text{p}})
\exp(-\tfrac{1}{4} w_{\text{p}}^2 |\mathbf{K}|^2) ,
\label{pompprof}
\end{equation}
where $w_{\text{p}}$ is the beam waist radius, $c$ is the speed of light, $\mathbf{K}$ is the two-dimensional transverse part of $\mathbf{k}$, $S(\omega;\delta\omega)$ is a normalized real-valued spectral function with a bandwidth of $\delta\omega$, and $\omega_{\text{p}}$ is the center frequency of the spectrum. The magnitude of the pump's angular spectrum is $\|\zeta(\mathbf{k})\|^2=|\zeta_0|^2$. The spectral function is normalized according to
\begin{equation}
\int S^2(\omega-\omega_{\text{p}};\delta\omega_{\text{p}})\ \frac{\text{d}\omega}{2\pi} = 1 .
\end{equation}
Under the monochromatic approximation, $S(\omega-\omega_{\text{p}};\delta\omega_{\text{p}})$ is a narrow function centered at $\omega_{\text{p}}$. For the calculations, we model the spectral function as a Gaussian.

For the parameter function in Eq.~(\ref{pompprof}), the bilinear vertex kernel under collinear type I phase matching conditions becomes
\begin{align}
H(\mathbf{k}_1,\mathbf{k}_2)
 = & i K_0(\omega_1,\omega_2) S(\omega_{\text{p}}-\omega_1-\omega_2;\delta\omega_{\text{p}}) \nonumber \\
& \times \exp\left(-\frac{w_{\text{p}}^2}{4 n_{\text{p}}^2} |n_1\mathbf{K}_1+n_2\mathbf{K}_2|^2 \right. \nonumber \\
& \left. +\frac{i 3 L c n_1 n_2
\left|\omega_2 \mathbf{K}_1 - \omega_1 \mathbf{K}_2\right|^2}{2 n_{\text{p}}\omega_{\text{p}} \omega_1 \omega_2}\right) ,
\label{skhaa}
\end{align}
where $n_{\text{p}}\equiv n_{\text{eff}}(\omega_{\text{p}})$, $n_1\equiv n_{\text{o}}(\omega_1)$, $n_2\equiv n_{\text{o}}(\omega_2)$, and
\begin{equation}
K_0(\omega_1,\omega_2) = \frac{4 \zeta_0^* \sigma_{\text{ooe}} w_{\text{p}} \omega_1 \omega_2
\sqrt{2\pi\omega_{\text{p}}}}{c^3 n_{\text{p}}^2} ,
\label{defK0}
\end{equation}
with $\sigma_{\text{ooe}}$ being the coefficient for the down-conversion process, represented as a scattering cross-section area.

\section{Thin crystal approximation}

The calculation can be alleviated by employing the {\em thin crystal approximation}, where the Rayleigh range of the pump beam is much longer than the length of the nonlinear crystal
\begin{equation}
L \ll \frac{w_{\text{p}}^2 \omega_{\text{p}}}{2c} .
\label{tcl}
\end{equation}
Such experimental conditions are valid in most parametric down-conversion experiments, especially those where the contribution of the spatial degrees of freedom are maximized \cite{law}.

The thin crystal approximation is to be separated from the down-conversion efficiency. Therefore, the $L$'s that appear in the exponents of the kernels contribute to the thin crystal approximation, while those outside the exponential functions combine into the down-conversion efficiency and do not play a role in the thin crystal approximation. If we remove the $L$-dependent part in the exponent of Eq.~(\ref{skhaa}), which means that we only consider the leading order term in the thin crystal approximation, then the computation of the higher order terms in the Magnus expansions of the semi-classical kernels simplify. Without the higher order thin crystal contributions in the exponents, the integrals can in some cases become singular. However, in the current analysis, where we overlap the kernels with the parameter function of the local oscillator, such situations do not occur.

The $z$ integrals in the Magnus expansion terms produce factors of $L^n/n!$ for contractions of $n$ bilinear kernels. Hence, the semi-classical kernels can be expressed as
\begin{align}
\begin{split}
A = & \mathbf{1} + \sum_{m=1}^{\infty} \frac{L^{2m}}{(2m)!} H_0^{\diamond 2m} \equiv \cosh_{\diamond}(LH_0) , \\
B = & i \sum_{m=1}^{\infty} \frac{L^{2m-1}}{(2m-1)!} H_0^{\diamond (2m-1)} \equiv i \sinh_{\diamond}(LH_0) ,
\end{split}
\label{loab}
\end{align}
where we set $H_{L=0}\equiv i H_0$. The subscript $\diamond$ indicates that all products in the expansions of these functions are represented by $\diamond$-contractions. Moreover, the first term in the expansion of $\cosh_{\diamond}(\cdot)$ is $\mathbf{1}$.

The integrals associated with the $\diamond$-contractions can be evaluated. For even and odd orders, respectively with $2m$ and $2m-1$ contractions, we obtain
\begin{align}
\begin{split}
H_0^{\diamond m_{\text{e}}} = & \frac{M_{\text{e}} M_1^{m_{\text{e}}}}{m_{\text{e}}^{5/4}}
\exp\left(-\frac{1}{4} \frac{w_{\text{p}}^2 n_1 |\mathbf{K}_1-\mathbf{K}_2|^2}{m_{\text{e}} n_{\text{p}}} \right) \\
 & \times S(\omega_1-\omega_2;\sqrt{m_{\text{e}}}\delta\omega_{\text{p}}) , \\
H_0^{\diamond m_{\text{o}}} = & \frac{M_{\text{o}} M_1^{m_{\text{o}}}}{m_{\text{o}}^{5/4}}
 \exp\left(-\frac{1}{4} \frac{w_{\text{p}}^2 |n_1\mathbf{K}_1+n_2\mathbf{K}_2|^2}{m_{\text{o}} n_{\text{p}}^2} \right) \\
 & \times S(\omega_{\text{p}}-\omega_1-\omega_2;\sqrt{m_{\text{o}}}\delta\omega_{\text{p}}) ,
 \end{split}
\label{heo}
\end{align}
where $m_{\text{e}}\equiv 2m$, $m_{\text{o}}\equiv 2m-1$, and
\begin{align}
\begin{split}
M_{\text{e}} = & \frac{\pi^{5/4} w_{\text{p}}^2 n_1^2 \omega_1}{c n_{\text{p}}^2\sqrt{\delta\omega_{\text{p}}}} , \\
M_{\text{o}} = & \frac{\pi^{5/4} w_{\text{p}}^2 n_1 n_2 \sqrt{\omega_1\omega_2}}{c n_{\text{p}}^2 \sqrt{\delta\omega_{\text{p}}}}
  , \\
M_1 = & \frac{4|\zeta_0|\sigma_{\text{ooe}}\sqrt{2\omega_{\text{p}}\omega_1\omega_2\delta\omega_{\text{p}}}}
{\pi^{3/4} w_{\text{p}} c^2 n_1 n_2} .
 \end{split}
\label{mmdefs}
\end{align}
Substituted into Eq.~(\ref{defzab}), these results lead to explicit expressions for the semi-classical kernels to all orders in the Magnus expansion.

The factors of $1/2m$ and $1/(2m-1)$ that appear in the exponents in Eq.~(\ref{heo}) have a significant effect. They cause these functions to become broader for larger $m$. A similar broadening effect is produced by the factors of $\sqrt{2m}$ and $\sqrt{2m-1}$ that appear with the bandwidths in the spectral functions. These factors appear as a result of the repeated convolutions involved in the multiple contractions and they only become apparent when considering the spatiotemporal degrees of freedom for all the terms in the Magnus expansion. The broadening plays an important role in the parametric process, as we'll see below.

\section{Quadrature variance}

Since the semi-classical approximation provides us with a Gaussian Wigner functional, the calculation of the variance in quadrature can be performed directly. These kernels commute and $A$ is real-valued. Therefore, we can substitute Eq.~(\ref{loab}) into Eq.~(\ref{qqverw}) to obtain
\begin{align}
\langle\hat{h}^2\rangle = & \gamma^*\diamond\cosh_{\diamond}(LH_0)\diamond\gamma
 -\tfrac{1}{2}\gamma\diamond\sinh_{\diamond}(LH_0)\diamond\gamma \nonumber \\
 & -\tfrac{1}{2}\gamma^*\diamond\sinh_{\diamond}(LH_0)\diamond\gamma^* .
\label{qqverw0}
\end{align}

First, we consider the case where the parameter function for the local oscillator is a Gaussian angular spectrum with a narrow temporal spectrum. It is given by
\begin{align}
\gamma(\mathbf{k}) = & \gamma_0 w_0 \sqrt{\frac{2\pi \omega}{c}} \Phi_0 S_0(\omega-\omega_{\text{d}};\delta\omega_{\text{d}})  \nonumber \\
 & \times \exp\left(-\tfrac{1}{4} w_0^2 |\mathbf{K}|^2 \right) ,
\label{ossdef}
\end{align}
where $w_0$ is the local oscillator beam waist radius, imaged to the crystal plane, $\Phi_0=\exp(i \theta)$, and $S_0(\omega-\omega_{\text{d}};\delta\omega_{\text{d}})$ is a narrow spectral function with a bandwidth $\delta\omega_{\text{d}}$, modeled as a Gaussian function centred at the degenerate down-conversion frequency $\omega_{\text{d}}$.

We substitute Eq.~(\ref{heo}) into Eq.~(\ref{loab}) and then, together with Eq.~(\ref{ossdef}), into Eq.~(\ref{qqverw}). After evaluating the integrals, we obtain the quadrature variance
\begin{align}
\sigma_q^2 = & \frac{\langle\hat{h}^2\rangle}{2\gamma_0^2} = \frac{1}{2}
+ \sum_{m=1}^{\infty} \frac{\Xi^{2m}(1+m\xi)^{-1/2}}{2(1+m\tau)(2m)!}  \nonumber \\
 & + \sin(2\theta) \sum_{m=1}^{\infty} \frac{\Xi^{2m-1}[1+\tfrac{1}{2}(2m-1)\xi]^{-1/2}}{[2+(2m-1)\tau](2m-1)!} , %\sqrt{2+(2m-1)\xi}
\end{align}
where
\begin{equation}
\xi \equiv \frac{\delta\omega_{\text{p}}^2}{\delta\omega_{\text{d}}^2} ~~~~~
\tau \equiv \frac{n_{\text{p}}^2 w_0^2}{n_{\text{d}}^2 w_{\text{p}}^2} ,
\label{grootrho}
\end{equation}
with $n_{\text{d}}\equiv n_{\text{o}}(\omega_{\text{d}})$, and
\begin{equation}
\Xi \equiv
\frac{2^{7/2}\pi^{5/4} L|\zeta_0|\sigma_{\text{ooe}} \sqrt{\delta_{\text{p}}}}{ w_{\text{p}} n_{\text{d}}^2\lambda_{\text{p}}^2} ,
\label{xidef}
\end{equation}
represents the efficiency of the down-conversion process, with the fractional bandwidth defined by
\begin{equation}
\delta_{\text{p}} \equiv \frac{\delta\lambda_{\text{p}}}{\lambda_{\text{p}}} = \frac{\delta\omega_{\text{p}}}{\omega_{\text{p}}} .
\end{equation}

\subsection{Small $\xi$ and small $\tau$}

The standard result of squeezing behavior is obtained when we assume that $w_{\text{p}}\gg w_0$ and $\delta\omega_{\text{d}}\gg \delta\omega_{\text{p}}$, so that $\xi\rightarrow 0$ and $\tau\rightarrow 0$. These conditions reproduce the well-known result
\begin{align}
\sigma_q^2 = & \frac{1}{2}
+ \frac{1}{2} \sum_{m=1}^{\infty} \frac{\Xi^{2m}}{(2m)!}
+ \frac{1}{2}\sin(2\theta)\sum_{m=1}^{\infty} \frac{\Xi^{2m-1}}{(2m-1)!} \nonumber \\
 = & \frac{1}{2} \cosh(\Xi) + \frac{1}{2} \sin(2\theta) \sinh(\Xi) .
\end{align}
The phase of the local oscillator is used to select two special orientations in phase space. For $\sin(2\theta)=\pm 1$,
\begin{equation}
\sigma_{\pm}^2 = \frac{1}{2} \exp(\pm\Xi) .
\end{equation}
These orientations represent those along which the Wigner functional has is maximum and minimum variance, respectively --- i.e., the squeezing direction and the direction orthogonal to it. The down-conversion efficiency $\Xi$, given in Eq.~(\ref{xidef}), is identified as the squeezing parameter. The product of the standard deviations along these orientations is a constant
\begin{equation}
\sigma_{+} \sigma_{-} = \frac{1}{2} ,
\end{equation}
which indicates that the state remains a minimum uncertainty state, regardless of the amount of squeezing.

\begin{figure}[ht]
\centerline{\includegraphics{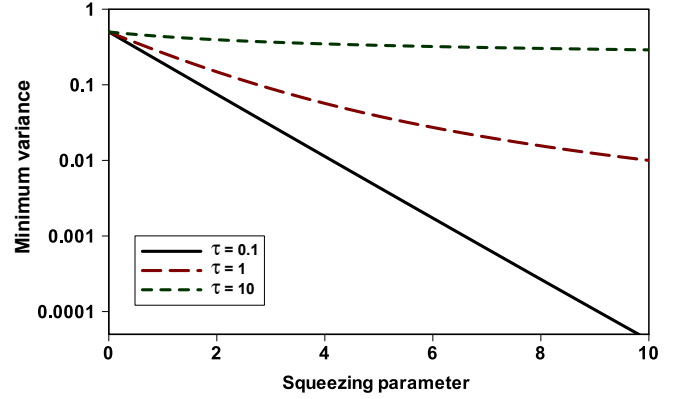}}
\caption{The minimum variance $\sigma_{-}^2$ is plotted as a function of the squeezing parameter $\Xi$ for $\tau=0.1$, $\tau=1$, and $\tau=10$.}
\label{wydte}
\end{figure}

In a practical situation, $\xi$ and $\tau$ will have finite values. Even if they are small, there will be a point where $m$ becomes large enough so that the approximations $\xi\rightarrow 0$ and $\tau\rightarrow 0$ are not valid anymore. Next, we'll consider the effect of these two cases in turn.

\begin{figure}[ht]
\centerline{\includegraphics{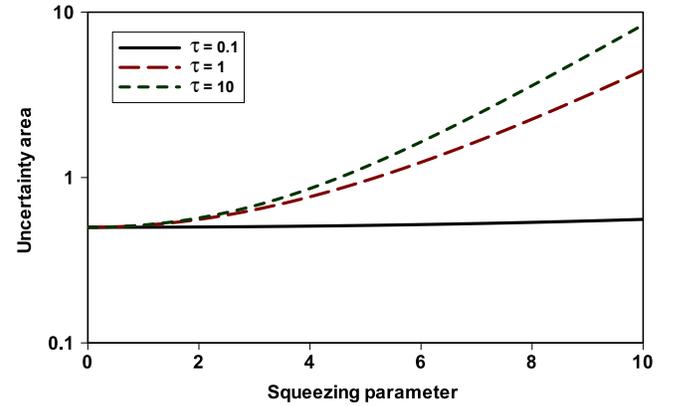}}
\caption{The uncertainty area size $\sigma_{+} \sigma_{-}$ as a function of the squeezing parameter $\Xi$ for $\tau=0.1$, $\tau=1$, and $\tau=10$.}
\label{area}
\end{figure}

\subsection{Small $\xi$ and arbitrary $\tau$}

Here we still assume $\xi\rightarrow 0$, but consider arbitrary $\tau$. For $\xi=0$, $\tau>0$ and $\sin(2\theta)=\pm 1$, the variance becomes
\begin{equation}
\sigma_{\pm}^2 = \sum_{n=0}^{\infty} \frac{(\pm\Xi)^n}{(n\tau+2)n!}
 = \frac{1}{2} {_1F_1}\left(\frac{2}{\tau};1+\frac{2}{\tau};\pm\Xi\right) ,
\label{qqreken2}
\end{equation}
where ${_1F_1}$ is a hypergeometric function. The result is that the squeezing effect is reduced. In Fig.~\ref{wydte}, the curves for minimum variance $\sigma_{-}^2$ are shown as a function of the squeezing parameter $\Xi$ for different values of $\tau$. It demonstrates the increase in the minimum variance as the value of $\tau$ increases.

The size of the uncertainty area of the state is shown in Fig.~\ref{area}, as a function of the squeezing parameter $\Xi$ for the same values of $\tau$ shown in Fig.~\ref{wydte}. We see that, for $\tau>0$, the state loses its minimum uncertainty property:
\begin{equation}
\sigma_{+} \sigma_{-} > \frac{1}{2} .
\end{equation}

\begin{figure}[ht]
\centerline{\includegraphics{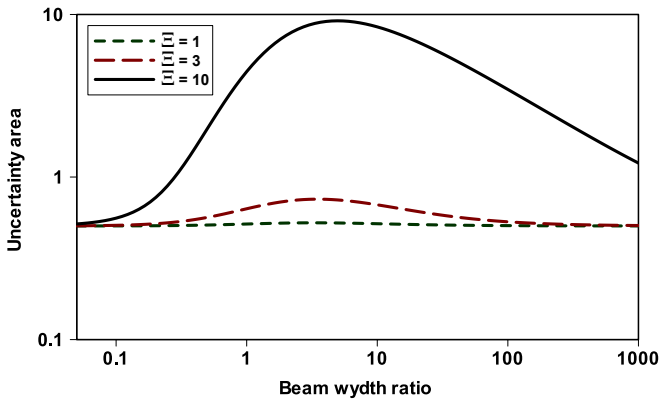}}
\caption{The uncertainty area size $\sigma_{+} \sigma_{-}$ as a function of the beam width ratio $\tau$ for $\Xi=1$, $\Xi=3$, and $\Xi=10$.}
\label{aparm}
\end{figure}

It is interesting to note that the increase in the size of the uncertainty area is not a monotonous function of $\tau$. In Fig.~\ref{aparm}, the uncertainty area is shown as a function of $\tau$ for different values of the squeezing parameter $\Xi$. It shows that the size of the uncertainty area peaks around $\tau\approx 6$ and then decreases again.

\subsection{Small $\tau$ and arbitrary $\xi$}

For the opposite situation, we assume $\tau\rightarrow 0$, and consider arbitrary $\xi$. For $\tau=0$, $\xi>0$, and $\sin(2\theta)=\pm 1$, we obtain a result that cannot be summed:
\begin{equation}
\sigma_{\pm}^2 = \sum_{n=0}^{\infty} \frac{(\pm\Xi)^n}{n!\sqrt{2}\sqrt{2+n\xi}} .
\end{equation}
However, with the aid of the auxiliary integral,
\begin{equation}
\frac{1}{\sqrt{a}} = \frac{1}{\sqrt{\pi}} \int \exp(-a x^2)\ dx ,
\end{equation}
we can represent it as an integral, given by
\begin{equation}
\sigma_{\pm}^2 = \int \frac{\exp(-2x^2)}{\sqrt{2\pi}}\exp\left[\pm\Xi\exp(-\xi x^2)\right]\ dx .
\end{equation}
It can be seen as an ensemble average of different curves with squeezing parameters varying as $\Xi\exp(-\xi x^2)$ for a Gaussian probability density. The effect of the ensemble averaging is to diminish the squeezing effect, depending on the value of $\xi$. If $\xi\ll 2$, the ensemble averaging effect is negligible, and one can set $\xi=0$.

In the case where both $\xi$ and $\tau$ can have arbitrary values, one would have a combination of these two effects. Hence, the optimal experimental conditions are those where $w_{\text{p}}\gg w_0$ and $\delta\omega_{\text{d}}\gg \delta\omega_{\text{p}}$.

\section{Laguerre-Gauss local oscillator}

Next, we consider the effect of using the Laguerre-Gauss modal basis for the local oscillator profile. Since the down-converted photon pairs under collinear phase matching conditions maintain the angular momentum of the pump photons that produced them, the azimuthal index of the one photon would always be the opposite of the other photon, if the pump has a zero azimuthal index. As a result, a local oscillator profile consisting of only one Laguerre-Gauss mode would only be able to see one of the two photons in a down-converted pair. Both photons are necessary for the squeezing effect. Therefore, such a single Laguerre-Gauss mode local oscillator would not see any squeezing.

For this reason, we consider petal modes that are formed as the superposition of two Laguerre-Gauss modes with opposite azimuthal indices. In terms of their angular spectra, the superposition is given by
\begin{equation}
\phi_{|\ell|,p}^{\text{(petal)}}(\mathbf{k}) = \frac{1}{\sqrt{2}}\phi_{\ell,p}^{\text{(LG)}}(\mathbf{k})
+ \frac{1}{\sqrt{2}}\phi_{-\ell,p}^{\text{(LG)}}(\mathbf{k}) .
\end{equation}
Note that a relative phase between the two Laguerre-Gauss modes would only produce an ineffectual rotation of the petal mode. Therefore, we do consider such a relative phase.

To alleviate the calculation, we use a generating function for the angular spectra of the Laguerre-Gauss modes, given by \cite{ipe,numipe}
\begin{align}
\mathcal{G}(\mathbf{k},\mu,\nu,s) = & \frac{w_0 \mathcal{N}_{\ell,p}}{1+\nu}
\exp\left[ \frac{i w_0(k_x+isk_y)\mu}{2(1+\nu)} \right] \nonumber \\
 & \times \exp\left[ -\frac{w_0^2(k_x^2+k_y^2)(1-\nu)}{4(1+\nu)} \right] ,
\end{align}
where $\mu$ and $\nu$ are generating parameters for the azimuthal and radial indices $\ell$ and $p$, respectively, and $s$ is a sign that represents the sign of the azimuthal index. The normalization constant is given by
\begin{equation}
\mathcal{N}_{\ell,p} = \sqrt{\frac{2^{|\ell|-1}p!}{\pi(p+|\ell|)!}} .
\end{equation}

The mode of the local oscillator is now defined in terms of the generating function:
\begin{align}
\gamma(\mathbf{k},\mu,\nu,s) = & 2\pi \Phi_0 \gamma_0 w_0 \sqrt{\frac{\omega}{c}} S_0(\omega-\omega_{\text{d}};\delta\omega_{\text{d}})  \nonumber \\
 & \times \mathcal{G}(\mathbf{k},\mu,\nu,s) .
\label{osslgdef}
\end{align}
The construction of the petal modes from this generating function is left till after the calculations.

The expression in Eq.~(\ref{osslgdef}) is substituted into Eq.~(\ref{qqverw0}), together with the expansions of the kernels in Eq.~(\ref{heo}). Since the generating function in Eq.~(\ref{osslgdef}) appears twice in each term, two different sets of generating parameters need to be used: $\{\mu_1,\nu_1,s_1\}$ and $\{\mu_2,\nu_2,s_2\}$. After evaluating all the integrals, we obtain
\begin{align}
\mathcal{J} = & 2\pi\mathcal{N}_{\ell,p} \sum_{m=0}^{\infty}
\left\{ \frac{\Xi^{2m} (1+m\xi)^{-1/2}}{d_{\text{e}}(\nu_1,\nu_1,\tau) (2m)!} \right. \nonumber \\
& \times \exp\left[\frac{\mu_1\mu_2(1+s_1 s_2)}{2 d_{\text{e}}(\nu_1,\nu_1,\tau)}\right] \nonumber \\
& \pm \frac{\Xi^{2m+1} [1+\tfrac{1}{2}(2m+1)\xi]^{-1/2}}{d_{\text{o}}(\nu_1,\nu_1,\tau) (2m+1)!} \nonumber \\
 & \left. \times \exp\left[\frac{\mu_1\mu_2(1-s_1 s_2)}{2 d_{\text{o}}(\nu_1,\nu_1,\tau)}\right] \right\} ,
\label{qqlgreken}
\end{align}
where
\begin{align}
\begin{split}
d_{\text{e}}(\nu_1,\nu_1,\tau) \equiv & 2m(1-\nu_1)(1-\nu_2)\tau + 2(1-\nu_1\nu_2) , \\
d_{\text{o}}(\nu_1,\nu_1,\tau) \equiv & (2m+1)(1-\nu_1)(1-\nu_2)\tau \\
& + 2(1-\nu_1\nu_2).
\end{split}
\end{align}

If the local oscillator consists of one Laguerre-Gauss mode only, we must set $s_1=s_2$, which implies that the second term in Eq.~(\ref{qqlgreken}) becomes independent of $\mu_1$ and $\mu_2$. Therefore, only the first term survives for $|\ell|>0$, leading to no squeezing.

At this point, we produce the petal modes by producing superpositions for both $s_1=\pm 1$ and $s_2=\pm 1$. The result then becomes
\begin{align}
\mathcal{J}' = &
\sum_{m=0}^{\infty} \frac{2\pi\mathcal{N}_{\ell,p} (\pm\Xi)^{m} (1+\tfrac{1}{2}m\xi)^{-1/2}}
{[m(1-\nu_1)(1-\nu_2)\tau + 2(1-\nu_1\nu_2)] m!} \nonumber \\
& \times \exp\left[\frac{\mu_1\mu_2}{m(1-\nu_1)(1-\nu_2)\tau + 2(1-\nu_1\nu_2)}\right] ,
\end{align}

If we assume $\xi\rightarrow 0$ and $\tau\rightarrow 0$, the expression reverts back to that for ideal squeezing, independent of $\ell$ and $p$. Therefore, under such conditions, the petal basis still supports squeezing in the same why one finds for $\ell=0$.

We will assume $\xi\rightarrow 0$, but allow arbitrary values for $\tau$. We also set the radial index to zero $p=0$, which is equivalent to setting $\nu_1=\nu_2=0$. The resulting expression for the generating function can then be used to generate the variance for arbitrary $|\ell|$:
\begin{equation}
\sigma_{\pm}^2 = \sum_{n=0}^{\infty} \frac{(\pm\Xi)^n 2^{|\ell|}}{(n\tau+2)^{|\ell|+1} n!} .
\end{equation}
It reproduces Eq.~(\ref{qqreken2}) for $\ell=0$, as expected. After evaluating the summation, we obtain the hypergeometric function
\begin{equation}
\sigma_{\pm}^2 = \tfrac{1}{2} {_tF_t}\left(M,..., M; N,..., N; \pm\Xi\right) ,
\end{equation}
where $t=|\ell|+1$, $M=2/\tau$, and $N=1+2/\tau$. The sequences of $M$'s and $N$'s in the argument of the hypergeometric function denote $|\ell|+1$ entries each.

\begin{figure}[ht]
\centerline{\includegraphics{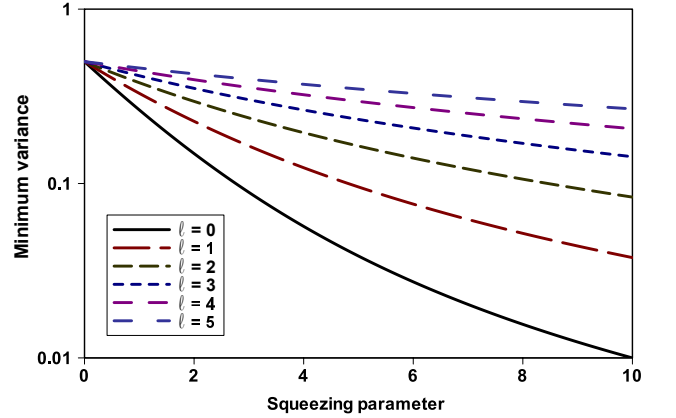}}
\caption{The minimum variance $\sigma_{-}^2$ is plotted as a function of the squeezing parameter $\Xi$ for $\tau=1$ and $\ell=0...5$.}
\label{lg}
\end{figure}

The minimum variance obtained when using a local oscillator with such petal mode profiles are shown in Fig.~\ref{lg} as a function of the squeezing parameter $\Xi$ for different values of the azimuthal index $\ell$ and with $\tau=1$. The rate of decrease in the minimum width slows down for larger azimuthal index.

\begin{figure}[ht]
\centerline{\includegraphics{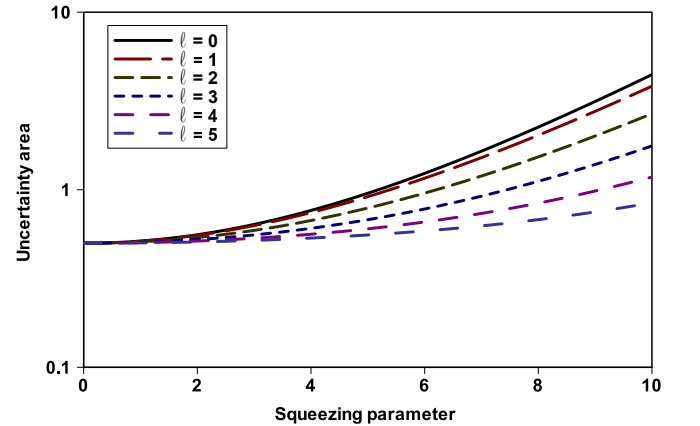}}
\caption{The uncertainty area size $\sigma_{+} \sigma_{-}$ as a function of the squeezing parameter $\Xi$ for $\tau=1$ and $\ell=0...5$.}
\label{motell}
\end{figure}

The curves of the uncertainty area for these petal mode profiles are shown in Fig.~\ref{motell} as a function of the squeezing parameter $\Xi$ for different values of the azimuthal index $\ell$ and with $\tau=1$. It is surprising to see that the uncertainty area increases slower for higher azimuthal indices than for lower azimuthal indices.

\begin{figure}[ht]
\centerline{\includegraphics{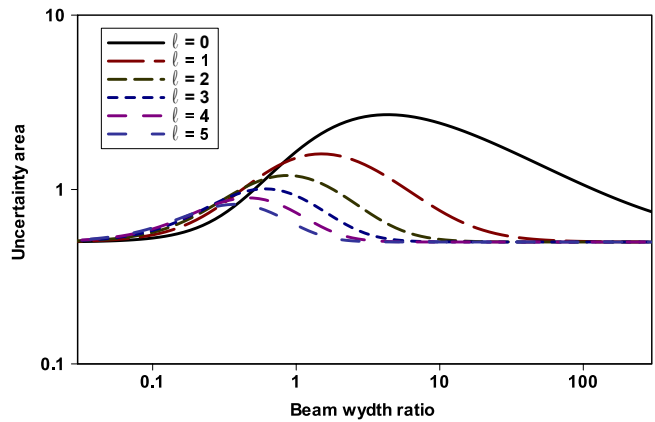}}
\caption{The uncertainty area size $\sigma_{+} \sigma_{-}$ as a function of the beam width ratio $\tau$ at $\Xi=7$ for $\ell=0...5$.}
\label{aparm2}
\end{figure}

We also consider the size of the uncertainty area as a function of the parameter $\tau$. In Fig.~\ref{aparm2}, it is shown for different values of the azimuthal index $\ell$ at $\Xi=7$. Again, we note that the size of the uncertainty area reaches a peak for a certain value of $\tau$ that depends on the  azimuthal index $\ell$ and then decreases for larger values of $\tau$.

\section{Discussion}

The effect of selected spatiotemporal properties of the local oscillator on the measurement of squeezing on a parametric down-converted state is investigated. The calculations are done directly in terms of the kernel functions and not on matrices derived from these kernel functions, which would require knowledge of the eigenbasis of the process. It makes it possible to perform calculations to all orders in the Magnus expansion of the down-converted state under thin crystal conditions.

The calculation demonstrates the efficacy of the Wigner functional approach: it incorporates all the spatiotemporal degrees of freedom without the need for discretization, which would require truncation to obtain explicit results. Yet, it contains all the relevant experimental parameters to make the analytically calculated result suitable for comparison with experimental results.

The Wigner functional approach analysis shows that there is a progressive broadening of the angular spectra and spectral functions for higher order terms in the Magnus expansion. The observed effect of the broadening depends on the experimental parameters of the local oscillator used for homodyne measurements of the amount of squeezing.

When the spatiotemporal degrees of freedom are ignored in analyses of the down-conversion process, one obtains the ideal squeezing behaviour. Such ideal behavior is also seen in analyses that incorporate the spatiotemporal degrees of freedom, but only goes as far as the first order term in the Magnus expansion \cite{horoshko}. It is only analyses that consider higher order terms in the Magnus expansion that would reveal the broadening in the widths of the angular spectra and spectral functions, leading to a reduction in the squeezing effect.

We investigate the parameter dependence of the reduction in squeezing, which deviates from the ideal squeezing behaviour. It is found that, to get as close to the ideal squeezing behaviour as possible, the mode size of the local oscillator should be as small as possible compared to the mode size of the pump and the bandwidth of the local oscillator should be much larger than that of the pump.

We also consider the effect of the azimuthal index when the local oscillator has a petal mode structure composed of the superposition of the Laguerre-Gauss mode with opposite azimuthal index. It is found that for a finite beam width ratio, the best squeezing behavior is obtained when the azimuthal index is zero.

Some interesting observations emerged from this investigation. The uncertainty area of the squeezed vacuum state at first increases as the beam width ratio increases, but then it starts to decrease again for further increases in the beam width ratio. We also found that, although the minimum variance is better for smaller azimuthal index, the size of the uncertainty area increases faster for smaller azimuthal index.

\section*{Acknowledgement}

This work was supported in part by funding from the National Research Foundation of South Africa (Grant Numbers: 118532).

% -------------

%\bibliographystyle{apsrev-title}
%\bibliography{fsr}

\begin{thebibliography}{33}
\expandafter\ifx\csname natexlab\endcsname\relax\def\natexlab#1{#1}\fi
\expandafter\ifx\csname bibnamefont\endcsname\relax
  \def\bibnamefont#1{#1}\fi
\expandafter\ifx\csname bibfnamefont\endcsname\relax
  \def\bibfnamefont#1{#1}\fi
\expandafter\ifx\csname citenamefont\endcsname\relax
  \def\citenamefont#1{#1}\fi
\expandafter\ifx\csname url\endcsname\relax
  \def\url#1{\texttt{#1}}\fi
\expandafter\ifx\csname urlprefix\endcsname\relax\def\urlprefix{URL }\fi
\providecommand{\bibinfo}[2]{#2}
\providecommand{\eprint}[2][]{\url{#2}}

\bibitem[{\citenamefont{Loudon and Knight}(1987)}]{sqlight}
\bibinfo{author}{\bibfnamefont{R.}~\bibnamefont{Loudon}} \bibnamefont{and}
  \bibinfo{author}{\bibfnamefont{P.~L.} \bibnamefont{Knight}}, ``Squeezed
  light,'' \bibinfo{journal}{J. Mod. Opt.} \textbf{\bibinfo{volume}{34}},
  \bibinfo{pages}{709} (\bibinfo{year}{1987}).

\bibitem[{\citenamefont{Lvovsky}(2015)}]{lvovsqu}
\bibinfo{author}{\bibfnamefont{A.~I.} \bibnamefont{Lvovsky}}, ``Squeezed
  light,'' \bibinfo{journal}{Photonics: Scientific Foundations, Technology and
  Applications} \textbf{\bibinfo{volume}{1}}, \bibinfo{pages}{121}
  (\bibinfo{year}{2015}).

\bibitem[{\citenamefont{Kolobov}(1999)}]{kolobov}
\bibinfo{author}{\bibfnamefont{M.~I.} \bibnamefont{Kolobov}}, ``The spatial
  behavior of nonclassical light,'' \bibinfo{journal}{Rev. Mod. Phys.}
  \textbf{\bibinfo{volume}{71}}, \bibinfo{pages}{1539} (\bibinfo{year}{1999}).

\bibitem[{\citenamefont{Braunstein}(2005)}]{braunstein}
\bibinfo{author}{\bibfnamefont{S.~L.} \bibnamefont{Braunstein}}, ``Squeezing as
  an irreducible resource,'' \bibinfo{journal}{Phys. Rev. A}
  \textbf{\bibinfo{volume}{71}}, \bibinfo{pages}{055801}
  (\bibinfo{year}{2005}).

\bibitem[{\citenamefont{Weedbrook et~al.}(2012)\citenamefont{Weedbrook,
  Pirandola, Garc{\'\i}a-Patr{\'o}n, Cerf, Ralph, Shapiro, and
  Lloyd}}]{weedbrook}
\bibinfo{author}{\bibfnamefont{C.}~\bibnamefont{Weedbrook}},
  \bibinfo{author}{\bibfnamefont{S.}~\bibnamefont{Pirandola}},
  \bibinfo{author}{\bibfnamefont{R.}~\bibnamefont{Garc{\'\i}a-Patr{\'o}n}},
  \bibinfo{author}{\bibfnamefont{N.~J.} \bibnamefont{Cerf}},
  \bibinfo{author}{\bibfnamefont{T.~C.} \bibnamefont{Ralph}},
  \bibinfo{author}{\bibfnamefont{J.~H.} \bibnamefont{Shapiro}},
  \bibnamefont{and} \bibinfo{author}{\bibfnamefont{S.}~\bibnamefont{Lloyd}},
  ``Gaussian quantum information,'' \bibinfo{journal}{Rev. Mod. Phys.}
  \textbf{\bibinfo{volume}{84}}, \bibinfo{pages}{621} (\bibinfo{year}{2012}).

\bibitem[{\citenamefont{Braunstein and Van~Loock}(2005)}]{contvar1}
\bibinfo{author}{\bibfnamefont{S.~L.} \bibnamefont{Braunstein}}
  \bibnamefont{and}
  \bibinfo{author}{\bibfnamefont{P.}~\bibnamefont{Van~Loock}}, ``Quantum
  information with continuous variables,'' \bibinfo{journal}{Rev. Mod. Phys.}
  \textbf{\bibinfo{volume}{77}}, \bibinfo{pages}{513} (\bibinfo{year}{2005}).

\bibitem[{\citenamefont{Adesso et~al.}(2014)\citenamefont{Adesso, Ragy, and
  Lee}}]{contvar2}
\bibinfo{author}{\bibfnamefont{G.}~\bibnamefont{Adesso}},
  \bibinfo{author}{\bibfnamefont{S.}~\bibnamefont{Ragy}}, \bibnamefont{and}
  \bibinfo{author}{\bibfnamefont{A.~R.} \bibnamefont{Lee}}, ``Continuous
  variable quantum information: Gaussian states and beyond,''
  \bibinfo{journal}{Open Syst. Inf. Dyn.} \textbf{\bibinfo{volume}{21}},
  \bibinfo{pages}{1440001} (\bibinfo{year}{2014}).

\bibitem[{\citenamefont{Andersen et~al.}(2016)\citenamefont{Andersen, Gehring,
  Marquardt, and Leuchs}}]{squreview}
\bibinfo{author}{\bibfnamefont{U.~L.} \bibnamefont{Andersen}},
  \bibinfo{author}{\bibfnamefont{T.}~\bibnamefont{Gehring}},
  \bibinfo{author}{\bibfnamefont{C.}~\bibnamefont{Marquardt}},
  \bibnamefont{and} \bibinfo{author}{\bibfnamefont{G.}~\bibnamefont{Leuchs}},
  ``30 years of squeezed light generation,'' \bibinfo{journal}{Physica Scripta}
  \textbf{\bibinfo{volume}{91}}, \bibinfo{pages}{053001}
  (\bibinfo{year}{2016}).

\bibitem[{\citenamefont{Aasi et~al.}(2013)\citenamefont{Aasi, Abadie, Abbott,
  Abbott, Abbott, Abernathy, Adams, Adams, Addesso, Adhikari et~al.}}]{ligosqu}
\bibinfo{author}{\bibfnamefont{J.}~\bibnamefont{Aasi}},
  \bibinfo{author}{\bibfnamefont{J.}~\bibnamefont{Abadie}},
  \bibinfo{author}{\bibfnamefont{B.~P.} \bibnamefont{Abbott}},
  \bibinfo{author}{\bibfnamefont{R.}~\bibnamefont{Abbott}},
  \bibinfo{author}{\bibfnamefont{T.~D.} \bibnamefont{Abbott}},
  \bibinfo{author}{\bibfnamefont{M.~R.} \bibnamefont{Abernathy}},
  \bibinfo{author}{\bibfnamefont{C.}~\bibnamefont{Adams}},
  \bibinfo{author}{\bibfnamefont{T.}~\bibnamefont{Adams}},
  \bibinfo{author}{\bibfnamefont{P.}~\bibnamefont{Addesso}},
  \bibinfo{author}{\bibfnamefont{R.~X.} \bibnamefont{Adhikari}},
  \bibnamefont{et~al.}, ``Enhanced sensitivity of the {LIGO} gravitational wave
  detector by using squeezed states of light,'' \bibinfo{journal}{Nature
  Photon.} \textbf{\bibinfo{volume}{7}}, \bibinfo{pages}{613}
  (\bibinfo{year}{2013}).

\bibitem[{\citenamefont{Hong and Mandel}(1985)}]{mandelspdc}
\bibinfo{author}{\bibfnamefont{C.~K.} \bibnamefont{Hong}} \bibnamefont{and}
  \bibinfo{author}{\bibfnamefont{L.}~\bibnamefont{Mandel}}, ``Theory of
  parametric frequency down conversion of light,'' \bibinfo{journal}{Phys. Rev.
  A} \textbf{\bibinfo{volume}{31}}, \bibinfo{pages}{2409}
  (\bibinfo{year}{1985}).

\bibitem[{\citenamefont{Arnaut and Barbosa}(2000)}]{arnaut}
\bibinfo{author}{\bibfnamefont{H.~H.} \bibnamefont{Arnaut}} \bibnamefont{and}
  \bibinfo{author}{\bibfnamefont{G.~A.} \bibnamefont{Barbosa}}, ``Orbital and
  intrinsic angular momentum of single photons and entangled pairs of photons
  generated by parametric down-conversion,'' \bibinfo{journal}{Phys. Rev.
  Lett.} \textbf{\bibinfo{volume}{85}}, \bibinfo{pages}{286}
  (\bibinfo{year}{2000}).

\bibitem[{\citenamefont{Bennink and Boyd}(2002)}]{bennink}
\bibinfo{author}{\bibfnamefont{R.~S.} \bibnamefont{Bennink}} \bibnamefont{and}
  \bibinfo{author}{\bibfnamefont{R.~W.} \bibnamefont{Boyd}}, ``Improved
  measurement of multimode squeezed light via an eigenmode approach,''
  \bibinfo{journal}{Phys. Rev. A} \textbf{\bibinfo{volume}{66}},
  \bibinfo{pages}{053815} (\bibinfo{year}{2002}).

\bibitem[{\citenamefont{Law and Eberly}(2004)}]{law}
\bibinfo{author}{\bibfnamefont{C.~K.} \bibnamefont{Law}} \bibnamefont{and}
  \bibinfo{author}{\bibfnamefont{J.~H.} \bibnamefont{Eberly}}, ``Analysis and
  interpretation of high transverse entanglement in optical parametric down
  conversion,'' \bibinfo{journal}{Phys. Rev. Lett.}
  \textbf{\bibinfo{volume}{92}}, \bibinfo{pages}{127903}
  (\bibinfo{year}{2004}).

\bibitem[{\citenamefont{Miatto et~al.}(2012)\citenamefont{Miatto, Pires,
  Barnett, and van Exter}}]{miatto3}
\bibinfo{author}{\bibfnamefont{F.~M.} \bibnamefont{Miatto}},
  \bibinfo{author}{\bibfnamefont{H.~D.~L.} \bibnamefont{Pires}},
  \bibinfo{author}{\bibfnamefont{S.~M.} \bibnamefont{Barnett}},
  \bibnamefont{and} \bibinfo{author}{\bibfnamefont{M.~P.} \bibnamefont{van
  Exter}}, ``Spatial {S}chmidt modes generated in parametric down-conversion,''
  \bibinfo{journal}{Eur. Phys. J. D} \textbf{\bibinfo{volume}{66}},
  \bibinfo{pages}{263} (\bibinfo{year}{2012}).

\bibitem[{\citenamefont{Gatti et~al.}(2012)\citenamefont{Gatti, Corti,
  Brambilla, and Horoshko}}]{gattispdc}
\bibinfo{author}{\bibfnamefont{A.}~\bibnamefont{Gatti}},
  \bibinfo{author}{\bibfnamefont{T.}~\bibnamefont{Corti}},
  \bibinfo{author}{\bibfnamefont{E.}~\bibnamefont{Brambilla}},
  \bibnamefont{and} \bibinfo{author}{\bibfnamefont{D.~B.}
  \bibnamefont{Horoshko}}, ``Dimensionality of the spatiotemporal entanglement
  of parametric down-conversion photon pairs,'' \bibinfo{journal}{Phys. Rev. A}
  \textbf{\bibinfo{volume}{86}}, \bibinfo{pages}{053803}
  (\bibinfo{year}{2012}).

\bibitem[{\citenamefont{Walborn et~al.}(2010)\citenamefont{Walborn, Monken,
  P{\'a}dua, and Souto~Ribeiro}}]{pdcspatcor}
\bibinfo{author}{\bibfnamefont{S.~P.} \bibnamefont{Walborn}},
  \bibinfo{author}{\bibfnamefont{C.~H.} \bibnamefont{Monken}},
  \bibinfo{author}{\bibfnamefont{S.}~\bibnamefont{P{\'a}dua}},
  \bibnamefont{and} \bibinfo{author}{\bibfnamefont{P.~H.}
  \bibnamefont{Souto~Ribeiro}}, ``Spatial correlations in parametric
  down-conversion,'' \bibinfo{journal}{Phys. Rep.}
  \textbf{\bibinfo{volume}{495}}, \bibinfo{pages}{87} (\bibinfo{year}{2010}).

\bibitem[{\citenamefont{Lanning et~al.}(2018)\citenamefont{Lanning, Xiao,
  Zhang, Novikova, Mikhailov, and Dowling}}]{oamsqu3}
\bibinfo{author}{\bibfnamefont{R.~N.} \bibnamefont{Lanning}},
  \bibinfo{author}{\bibfnamefont{Z.}~\bibnamefont{Xiao}},
  \bibinfo{author}{\bibfnamefont{M.}~\bibnamefont{Zhang}},
  \bibinfo{author}{\bibfnamefont{I.}~\bibnamefont{Novikova}},
  \bibinfo{author}{\bibfnamefont{E.~E.} \bibnamefont{Mikhailov}},
  \bibnamefont{and} \bibinfo{author}{\bibfnamefont{J.~P.}
  \bibnamefont{Dowling}}, ``Quantized nonlinear {G}aussian-beam dynamics:
  Tailoring multimode squeezed-light generation,'' \bibinfo{journal}{Phys. Rev.
  A} \textbf{\bibinfo{volume}{98}}, \bibinfo{pages}{043824}
  (\bibinfo{year}{2018}).

\bibitem[{\citenamefont{Horoshko et~al.}(2019)\citenamefont{Horoshko, La~Volpe,
  Arzani, Treps, Fabre, and Kolobov}}]{horoshko}
\bibinfo{author}{\bibfnamefont{D.~B.} \bibnamefont{Horoshko}},
  \bibinfo{author}{\bibfnamefont{L.}~\bibnamefont{La~Volpe}},
  \bibinfo{author}{\bibfnamefont{F.}~\bibnamefont{Arzani}},
  \bibinfo{author}{\bibfnamefont{N.}~\bibnamefont{Treps}},
  \bibinfo{author}{\bibfnamefont{C.}~\bibnamefont{Fabre}}, \bibnamefont{and}
  \bibinfo{author}{\bibfnamefont{M.~I.} \bibnamefont{Kolobov}},
  ``Bloch-{M}essiah reduction for twin beams of light,''
  \bibinfo{journal}{Phys. Rev. A} \textbf{\bibinfo{volume}{100}},
  \bibinfo{pages}{013837} (\bibinfo{year}{2019}).

\bibitem[{\citenamefont{Bloch and Messiah}(1962)}]{blochmessiah}
\bibinfo{author}{\bibfnamefont{C.}~\bibnamefont{Bloch}} \bibnamefont{and}
  \bibinfo{author}{\bibfnamefont{A.}~\bibnamefont{Messiah}}, ``The canonical
  form of an antisymmetric tensor and its application to the theory of
  superconductivity,'' \bibinfo{journal}{Nucl. Phys.}
  \textbf{\bibinfo{volume}{39}}, \bibinfo{pages}{95} (\bibinfo{year}{1962}).

\bibitem[{\citenamefont{Magnus}(1954)}]{magnus}
\bibinfo{author}{\bibfnamefont{W.}~\bibnamefont{Magnus}}, ``On the exponential
  solution of differential equations for a linear operator,''
  \bibinfo{journal}{Commun. Pure Appl. Math.} \textbf{\bibinfo{volume}{7}},
  \bibinfo{pages}{649} (\bibinfo{year}{1954}).

\bibitem[{\citenamefont{Blanes et~al.}(2009)\citenamefont{Blanes, Casas, Oteo,
  and Ros}}]{magnusrev}
\bibinfo{author}{\bibfnamefont{S.}~\bibnamefont{Blanes}},
  \bibinfo{author}{\bibfnamefont{F.}~\bibnamefont{Casas}},
  \bibinfo{author}{\bibfnamefont{J.~A.} \bibnamefont{Oteo}}, \bibnamefont{and}
  \bibinfo{author}{\bibfnamefont{J.}~\bibnamefont{Ros}}, ``The {M}agnus
  expansion and some of its applications,'' \bibinfo{journal}{Phys. Rep.}
  \textbf{\bibinfo{volume}{470}}, \bibinfo{pages}{151} (\bibinfo{year}{2009}).

\bibitem[{\citenamefont{Roux}(2020{\natexlab{a}})}]{nosemi}
\bibinfo{author}{\bibfnamefont{F.~S.} \bibnamefont{Roux}}, ``Parametric
  down-conversion beyond the semiclassical approximation,''
  \bibinfo{journal}{Phys. Rev. Research} \textbf{\bibinfo{volume}{2}},
  \bibinfo{pages}{033398} (\bibinfo{year}{2020}{\natexlab{a}}).

\bibitem[{\citenamefont{Roux}(2018)}]{stquad}
\bibinfo{author}{\bibfnamefont{F.~S.} \bibnamefont{Roux}}, ``Combining
  spatiotemporal and particle-number degrees of freedom,''
  \bibinfo{journal}{Phys. Rev. A} \textbf{\bibinfo{volume}{98}},
  \bibinfo{pages}{043841} (\bibinfo{year}{2018}).

\bibitem[{\citenamefont{Roux}(2020{\natexlab{b}})}]{stquaderr}
\bibinfo{author}{\bibfnamefont{F.~S.} \bibnamefont{Roux}}, ``Erratum:
  {C}ombining spatiotemporal and particle-number degrees of freedom [{P}hys.
  {R}ev. {A} 98, 043841 (2018)],'' \bibinfo{journal}{Phys. Rev. A}
  \textbf{\bibinfo{volume}{101}}, \bibinfo{pages}{019903(E)}
  (\bibinfo{year}{2020}{\natexlab{b}}).

\bibitem[{\citenamefont{Quesada and Sipe}(2014)}]{sipemagnus1}
\bibinfo{author}{\bibfnamefont{N.}~\bibnamefont{Quesada}} \bibnamefont{and}
  \bibinfo{author}{\bibfnamefont{J.~E.} \bibnamefont{Sipe}}, ``Effects of time
  ordering in quantum nonlinear optics,'' \bibinfo{journal}{Phys. Rev. A}
  \textbf{\bibinfo{volume}{90}}, \bibinfo{pages}{063840}
  (\bibinfo{year}{2014}).

\bibitem[{\citenamefont{Quesada and Sipe}(2015)}]{sipemagnus2}
\bibinfo{author}{\bibfnamefont{N.}~\bibnamefont{Quesada}} \bibnamefont{and}
  \bibinfo{author}{\bibfnamefont{J.~E.} \bibnamefont{Sipe}}, ``Time-ordering
  effects in the generation of entangled photons using nonlinear optical
  processes,'' \bibinfo{journal}{Phys. Rev. Lett.}
  \textbf{\bibinfo{volume}{114}}, \bibinfo{pages}{093903}
  (\bibinfo{year}{2015}).

\bibitem[{\citenamefont{Lassen et~al.}(2009)\citenamefont{Lassen, Leuchs, and
  Andersen}}]{oamsqu1}
\bibinfo{author}{\bibfnamefont{M.}~\bibnamefont{Lassen}},
  \bibinfo{author}{\bibfnamefont{G.}~\bibnamefont{Leuchs}}, \bibnamefont{and}
  \bibinfo{author}{\bibfnamefont{U.~L.} \bibnamefont{Andersen}}, ``Continuous
  variable entanglement and squeezing of orbital angular momentum states,''
  \bibinfo{journal}{Phys. Rev. Lett.} \textbf{\bibinfo{volume}{102}},
  \bibinfo{pages}{163602} (\bibinfo{year}{2009}).

\bibitem[{\citenamefont{Hsu et~al.}(2009)\citenamefont{Hsu, Bowen, and
  Lam}}]{oamsqu2}
\bibinfo{author}{\bibfnamefont{M.~T.~L.} \bibnamefont{Hsu}},
  \bibinfo{author}{\bibfnamefont{W.~P.} \bibnamefont{Bowen}}, \bibnamefont{and}
  \bibinfo{author}{\bibfnamefont{P.~K.} \bibnamefont{Lam}}, ``Spatial-state
  {S}tokes-operator squeezing and entanglement for optical beams,''
  \bibinfo{journal}{Phys. Rev. A} \textbf{\bibinfo{volume}{79}},
  \bibinfo{pages}{043825} (\bibinfo{year}{2009}).

\bibitem[{\citenamefont{Pecoraro et~al.}(2019)\citenamefont{Pecoraro, Cardano,
  Marrucci, and Porzio}}]{cvoamappl}
\bibinfo{author}{\bibfnamefont{A.}~\bibnamefont{Pecoraro}},
  \bibinfo{author}{\bibfnamefont{F.}~\bibnamefont{Cardano}},
  \bibinfo{author}{\bibfnamefont{L.}~\bibnamefont{Marrucci}}, \bibnamefont{and}
  \bibinfo{author}{\bibfnamefont{A.}~\bibnamefont{Porzio}},
  ``Continuous-variable entangled states of light carrying orbital angular
  momentum,'' \bibinfo{journal}{Phys. Rev. A} \textbf{\bibinfo{volume}{100}},
  \bibinfo{pages}{012321} (\bibinfo{year}{2019}).

\bibitem[{\citenamefont{Zhang et~al.}(2018)\citenamefont{Zhang, Jin, Zhang,
  Cen, Hu, and Zhao}}]{oamsquappl}
\bibinfo{author}{\bibfnamefont{J.-D.} \bibnamefont{Zhang}},
  \bibinfo{author}{\bibfnamefont{C.-F.} \bibnamefont{Jin}},
  \bibinfo{author}{\bibfnamefont{Z.-J.} \bibnamefont{Zhang}},
  \bibinfo{author}{\bibfnamefont{L.-Z.} \bibnamefont{Cen}},
  \bibinfo{author}{\bibfnamefont{J.-Y.} \bibnamefont{Hu}}, \bibnamefont{and}
  \bibinfo{author}{\bibfnamefont{Y.}~\bibnamefont{Zhao}}, ``Super-sensitive
  angular displacement estimation via an {SU}(1,1)-{SU}(2) hybrid
  interferometer,'' \bibinfo{journal}{Opt. Express}
  \textbf{\bibinfo{volume}{26}}, \bibinfo{pages}{33080} (\bibinfo{year}{2018}).

\bibitem[{\citenamefont{Roux}(2020{\natexlab{c}})}]{entpdc}
\bibinfo{author}{\bibfnamefont{F.~S.} \bibnamefont{Roux}}, ``Quantifying
  entanglement of parametric down-converted states in all degrees of freedom,''
  \bibinfo{journal}{Phys. Rev. Research} \textbf{\bibinfo{volume}{2}},
  \bibinfo{pages}{023137} (\bibinfo{year}{2020}{\natexlab{c}}).

\bibitem[{\citenamefont{Roux}(2011)}]{ipe}
\bibinfo{author}{\bibfnamefont{F.~S.} \bibnamefont{Roux}},
  ``Infinitesimal-propagation equation for decoherence of an
  orbital-angular-momentum-entangled biphoton state in atmospheric
  turbulence,'' \bibinfo{journal}{Phys. Rev. A} \textbf{\bibinfo{volume}{83}},
  \bibinfo{pages}{053822} (\bibinfo{year}{2011}).

\bibitem[{\citenamefont{Mabena and Roux}(2019)}]{numipe}
\bibinfo{author}{\bibfnamefont{C.~M.} \bibnamefont{Mabena}} \bibnamefont{and}
  \bibinfo{author}{\bibfnamefont{F.~S.} \bibnamefont{Roux}}, ``Optical orbital
  angular momentum under strong scintillation,'' \bibinfo{journal}{Phys. Rev.
  A} \textbf{\bibinfo{volume}{99}}, \bibinfo{pages}{013828}
  (\bibinfo{year}{2019}).

\end{thebibliography}

\end{document}